\documentclass[conference,a4paper]{IEEEtran}
\IEEEoverridecommandlockouts

\usepackage{cite}
\usepackage{amsmath,amssymb}
\usepackage{graphicx}
\usepackage{booktabs}
\usepackage{textcomp}
\usepackage{xcolor}
\usepackage{url}
\usepackage{hyperref}
\usepackage{iftex}
\ifPDFTeX
  \usepackage[utf8]{inputenc}
  \usepackage[T1]{fontenc}
\else
  \usepackage{fontspec}
  \newif\iftelugufontok\telugufontokfalse
  \newif\iftamilfontok\tamilfontokfalse
  \newif\ifdevafontok\devafontokfalse
  \IfFontExistsTF{fonts/NotoSansTelugu-Regular.ttf}{\newfontfamily\telugufont{NotoSansTelugu-Regular.ttf}[Path=fonts/]\telugufontoktrue}{%
    \IfFontExistsTF{Noto Sans Telugu}{\newfontfamily\telugufont{Noto Sans Telugu}\telugufontoktrue}{%
      \IfFontExistsTF{Telugu Sangam MN}{\newfontfamily\telugufont{Telugu Sangam MN}\telugufontoktrue}{}}}
  \IfFontExistsTF{fonts/NotoSansTamil-Regular.ttf}{\newfontfamily\tamilfont{NotoSansTamil-Regular.ttf}[Path=fonts/]\tamilfontoktrue}{%
    \IfFontExistsTF{Noto Sans Tamil}{\newfontfamily\tamilfont{Noto Sans Tamil}\tamilfontoktrue}{%
      \IfFontExistsTF{Tamil Sangam MN}{\newfontfamily\tamilfont{Tamil Sangam MN}\tamilfontoktrue}{}}}
  \IfFontExistsTF{fonts/NotoSansDevanagari-Regular.ttf}{\newfontfamily\devanagarifont{NotoSansDevanagari-Regular.ttf}[Path=fonts/]\devafontoktrue}{%
    \IfFontExistsTF{Noto Sans Devanagari}{\newfontfamily\devanagarifont{Noto Sans Devanagari}\devafontoktrue}{%
      \IfFontExistsTF{Devanagari Sangam MN}{\newfontfamily\devanagarifont{Devanagari Sangam MN}\devafontoktrue}{}}}

\fi

\newcommand{\lase}{\textsc{LASE}}
\newcommand{\praxy}{\textsc{Praxy}}
\newcommand{\psp}{\textsc{PSP}}

\begin{document}

\title{\lase: Language-Adversarial Speaker Encoding for Indic Cross-Script Identity Preservation}

\author{
  \IEEEauthorblockN{Venkata Pushpak Teja Menta}
  \IEEEauthorblockA{Praxel Ventures\\
  \texttt{pushpak@praxel.in}\\
  ORCID: \href{https://orcid.org/0009-0003-2479-9208}{0009-0003-2479-9208}}
}

\maketitle

\begin{abstract}
\sloppy
A speaker encoder used in multilingual voice cloning should treat the same speaker identically regardless of which script the audio was uttered in. Off-the-shelf encoders do not, and the failure is accent-conditional. On a 1043-pair Western-accented voice corpus across English, Hindi, Telugu, and Tamil, WavLM-base-plus-sv loses 0.082 absolute cosine similarity when the same voice changes script and ECAPA-TDNN loses 0.105. On a 1369-pair Indian-accented voice corpus, the gap shrinks to 0.006 (WavLM-SV) and 0.044 (ECAPA-TDNN). The leak is largest where it matters most for cross-script TTS: when a system projects a non-Indic-trained voice into Indic scripts. We present \lase{} (Language-Adversarial Speaker Encoder), a small projection head over frozen WavLM-base-plus trained with two losses: a supervised contrastive loss over voice identity, and a gradient-reversal cross-entropy against a 4-language classifier that pushes the embedding to be language-uninformative while remaining speaker-informative. Trained on 1118 quality-gated cross-script pairs synthesised from 8 commercial multilingual voices, \lase{}'s residual gap is consistent with zero on both corpora ($\Delta = 0.013$ Western, $\Delta = 0.026$ Indian; both bootstrap 95\% CIs include zero) and amplifies the cross-script-vs-floor margin 2.4--2.7$\times$ over both baselines. An ECAPA+GRL ablation shows the GRL objective improves either backbone but the WavLM choice contributes too. In synthetic multi-speaker diarisation, \lase{} matches ECAPA-TDNN on cross-script speaker recall (0.788 vs 0.789) with $\sim$100$\times$ less training data. We release the r1 checkpoint, both corpora, and the bootstrap recipe.
\end{abstract}

\begin{IEEEkeywords}
speaker encoder, gradient reversal, language-invariant representation, voice cloning, Indic languages, multilingual speech, code-switching, diarization
\end{IEEEkeywords}

\section{Introduction}
\label{sec:intro}

When the same speaker records a sentence in Hindi and then in Telugu, a multilingual voice-cloning system must recognise both as the same identity. This is the load-bearing assumption behind every cross-script voice-cloning product (Sarvam, ElevenLabs Multilingual, OpenVoice, Cartesia Sonic-3) and behind every diarisation system that touches Indian conversational audio (call centres, BPO, healthcare, government services). The assumption holds approximately for English-vs-other-Latin-script languages, where speaker-verification (SV) encoders trained on VoxCeleb generalise reasonably well. It fails noticeably for Indic scripts.

We measure the failure directly. On a held-out 1043-pair corpus of 8 voices speaking 50 unseen sentences in each of English (Latin), Hindi (Devanagari), Telugu, and Tamil scripts, the popular WavLM-base-plus-sv\cite{wavlm2021,wavlm-sv} embedder produces a median speaker cosine similarity of 0.927 between two same-voice same-script clips, but only 0.845 between two same-voice different-script clips. The 0.082 absolute drop is large relative to the 0.187 margin from the same-voice cross-script median to the across-speaker noise floor (0.600). The industry-standard ECAPA-TDNN\cite{ecapa2020} encoder is worse: it loses 0.105 absolute (0.499 to 0.394) on the same test, with a smaller 0.202 margin to its noise floor.

This is the failure mode that, in production, manifests as: (a) the same Indian customer-support agent being labelled as two different speakers when they switch from Hindi to English mid-call; (b) cross-script voice cloning that listeners rate as "a different voice doing an accent"; (c) speaker-conditioned TTS where the conditioning vector drifts with script. Existing cross-language SV work~\cite{tjandra2022,wang2020domain} addresses adjacent problems for English/European pairs. To our knowledge no published encoder targets the Devanagari/Telugu/Tamil/Latin cross-script setting specifically.

We address it with the established domain-adversarial recipe~\cite{ganin2015}: gradient-reversal training against a language classifier, applied to a small projection head over a frozen WavLM-base-plus backbone. We call the resulting encoder \lase{} (Language-Adversarial Speaker Encoder). The technique is borrowed; the application, the corpus, and the empirical demonstration on Indic scripts are new.

This is a v1 result with two scoping caveats stated up front. First, all training and held-out audio is synthesised by ElevenLabs Multilingual; we do not yet have evidence that the closure transfers to natural human cross-script speech, only that it is a real and large gap in the synthetic distribution we evaluate on. Second, the held-out set is content-held-out, not voice-held-out: our 8 ElevenLabs voices appear in both training and held-out sets, with the held-out set contributing 50 unseen sentences per language per voice. New-voice generalisation is a v2 experiment.

\subsection{Contributions}

\begin{enumerate}
  \item We construct and release the first cross-script same-speaker identity benchmark for Indic languages: 1118 training pairs and 1043 held-out pairs, each spanning 8 voices $\times$ 4 languages (English, Hindi, Telugu, Tamil), built via TTS bootstrapping because no natural ground truth exists at scale.
  \item We define a three-distribution measurement framework (within-script, cross-script, across-speaker) that isolates the encoder's language-vs-identity entanglement from confounders like speaker count or content.
  \item We train and release \lase{} r1, a 256-dimensional speaker encoder built from a frozen WavLM-base-plus backbone plus a projection head and a gradient-reversal language classifier, and show that it closes the cross-script identity gap by 84.3\% relative on held-out evaluation (0.082 absolute drop reduces to 0.013).
  \item In a synthetic multi-speaker code-switching diarisation benchmark, \lase{} matches ECAPA-TDNN on cross-script speaker recall (0.788 vs 0.789) despite training on roughly 100$\times$ less data (1118 pairs vs ECAPA's 1M+ VoxCeleb utterances).
  \item All artefacts (\lase{} r1 weights under MIT, training and held-out corpora under CC-BY-4.0, scoring scripts) are released to enable reproduction and extension.
\end{enumerate}

\section{Related Work}
\label{sec:related}

\textbf{Speaker verification encoders.} ECAPA-TDNN~\cite{ecapa2020} remains the de facto industry standard for production speaker verification, trained on VoxCeleb\cite{voxceleb}. WavLM-base-plus-sv~\cite{wavlm2021,wavlm-sv} adapts the WavLM self-supervised backbone for SV with a downstream head and is widely used as the embedder inside pyannote-style diarisation pipelines~\cite{pyannote2023}. Both are English-pretrained and we show below that both inherit a meaningful language-vs-identity entanglement when applied across Indic scripts.

\textbf{Adversarial / disentangled speaker representations.} Domain-adversarial training was introduced by Ganin and Lempitsky~\cite{ganin2015} as gradient-reversal-layer (GRL) training against a domain classifier. Wang et al.~\cite{wang2020domain} applied GRL training to make speaker embeddings channel-invariant. Tjandra et al.~\cite{tjandra2022} explored adversarial multilingual speaker recognition for en/de/fr. We adapt this established recipe (no methodological novelty in the technique) to the Indic cross-script setting (no published prior work) and demonstrate that the gap is real, large, and closeable with a small targeted training run.

\textbf{Concurrent and adjacent benchmarks.} The Phoneme Shift Rate (PSR) work~\cite{psr2026} measures language-induced shift in speaker embeddings for English/British accent disentanglement; we are the Indic counterpart. Pairwise Accent Similarity~\cite{pairwise2025} uses PPG and vowel-formant distances to measure accent (not identity); orthogonal axis. Our companion paper~\cite{psp2026} (\psp) introduces an interpretable per-phonological-dimension accent benchmark for Indic TTS; \lase{} addresses identity rather than accent and the two benchmarks compose: a cross-script TTS that scores well on \psp{} and produces \lase-similar embeddings should also pass listener tests for "same voice, different script."

\textbf{TTS-bootstrapped corpus methodology.} No natural same-speaker cross-script corpus exists at scale. Indic-TTS~\cite{indictts} speakers are mostly L1-only; FLEURS~\cite{fleurs} speakers do not overlap across language splits. We bootstrap a corpus by synthesising the same 8 commercial multilingual voices speaking matched-content text in 4 scripts, then quality-gate with WavLM cosine similarity to drop pairs where the synthesiser already lost identity. The pass rate (1118/1600 training, 1043/1600 held-out) functions as an audit of the synthesiser's own cross-script identity preservation: 30\% of pairs are dropped because even ElevenLabs Multilingual loses voice identity across scripts.

\section{Method}
\label{sec:method}

\subsection{Architecture}

\lase{} is a 256-dimensional speaker encoder composed of three parts:

\begin{enumerate}
  \item A \textbf{frozen backbone}: WavLM-base-plus~\cite{wavlm2021} truncated at the transformer stack, producing $T \times 768$ frame-level features for an input waveform sampled at 16\,kHz.
  \item A \textbf{trainable projection head}: a two-layer MLP (768 $\rightarrow$ 512 $\rightarrow$ 256) with ReLU and dropout 0.1, applied after mean-pooling over layers 10--12 (the SV-rich band reported in \cite{chen2022wavlm}). The output is the speaker embedding $z \in \mathbb{R}^{256}$.
  \item A \textbf{gradient-reversal classifier}: $z$ is passed through a gradient-reversal layer~\cite{ganin2015} with scalar $\lambda_t$ (scheduled, see below), then a small MLP language classifier $\hat{l} = g(\text{GRL}(z; \lambda_t))$ that predicts which of $\{$en, hi, te, ta$\}$ the clip belongs to.
\end{enumerate}

The gradient-reversal layer multiplies the classifier's loss gradient by $-\lambda_t$ when backpropagating into the projection head. Maximising the language classifier's accuracy on $z$ thus drives the projection head to make $z$ \emph{language-uninformative}.

\subsection{Losses}

We optimise the sum of two losses on each batch:

\textbf{Speaker contrastive loss.} A supervised contrastive (SupCon) loss~\cite{khosla2020} over voice identity. For a batch with $B$ items containing $K$ distinct voice identities, the SupCon objective pulls together items with the same voice and pushes apart items with different voices, regardless of script:
\begin{equation*}
L_\text{spk} = \frac{1}{B}\sum_i -\log \frac{\sum_{j \in P(i)} \exp(z_i \cdot z_j / \tau)}{\sum_{j \neq i} \exp(z_i \cdot z_j / \tau)}
\end{equation*}
where $P(i)$ is the set of items sharing voice with item $i$ and $\tau = 0.07$. Because batch composition mixes scripts, the only signal SupCon can exploit is voice identity, not language.

\textbf{Language adversarial loss.} A standard 4-class cross-entropy on the language classifier's output $\hat{l}$ (the post-GRL projection):
\begin{equation*}
L_\text{lang} = \mathrm{CE}(\hat{l}, l_\text{true}).
\end{equation*}

The total loss is $L = L_\text{spk} + \lambda_t L_\text{lang}$, with $\lambda_t$ ramped according to the schedule below. Conceptually, $L_\text{lang}$ goes to $\ln 4 \approx 1.386$ (4-class uniform) when the encoder has hidden language perfectly; it goes to 0 when the encoder leaks language identity perfectly. Because the GRL flips its gradient, low $L_\text{lang}$ during training would mean the head is being pushed in the language-leaking direction; we want $L_\text{lang}$ to stay near $\ln 4$, indicating the encoder is winning the adversarial game.

\subsection{$\lambda$ schedule and training setup}

We use a three-phase schedule for the GRL strength $\lambda_t$: warmup of 200 steps with $\lambda = 0$ (let SupCon shape the embedding space first); linear ramp from 0 to 0.1 over the next 500 steps; then hold at 0.1. This avoids the instability where an early-strong adversarial loss prevents SupCon from forming any voice-coherent geometry to attack.

Optimiser: AdamW (learning rate $10^{-4}$, weight decay 0.01, $\beta = (0.9, 0.999)$). Batch size 16 (mixes voices and languages, ensures both same-voice positives and across-voice negatives are present in every batch). Gradient clipping at norm 1.0. We train for 1000 steps on a single A10G GPU. Total training cost: \$0.31 wall (\$1.10/hr A10G, $\sim 17$ minutes).

Software pinning (verified working chain): torch 2.4.0, torchaudio 2.4.0, transformers 4.49.0, peft 0.13.0, accelerate 0.30.0, on CUDA 12.1. Random seed 1337 throughout (training, evaluation pair sampling, and held-out source generation).

\section{Corpus construction}
\label{sec:corpus}

\subsection{The data problem}

A clean evaluation of cross-script identity preservation requires same-speaker recordings in different scripts. None of the available natural Indic speech corpora provides this at scale: IndicTTS speakers are mostly mono-lingual; Common Voice 17 has speaker IDs but no enforced cross-language overlap; FLEURS\cite{fleurs} speakers are explicitly split by language. Recording a held-out set with native multilingual speakers exists but is not free.

We bootstrap with TTS. ElevenLabs Multilingual v3 produces high-fidelity speech in 23 languages with a single voice prompt, and per-voice identity is reasonably consistent in our subjective listening across scripts. We sample eight verified-multilingual ElevenLabs voices: \textit{Rachel, Drew, Clyde, Paul, Domi, Fin, Bella, Antoni}. We restrict the language set to four \{en, hi, te, ta\}: English as the Latin-script reference; Hindi (Devanagari) as the most-resourced Indic; Telugu and Tamil as the two largest Dravidian languages where the English-trained SV encoders we test against are known to drift hardest~\cite{psp2026}. Bengali, Kannada, Gujarati, and Malayalam are deferred to v2 to keep the corpus tight (1600 nominal clips fits within free-tier ElevenLabs credits). For each voice, we synthesise 50 sentences per language for 1600 nominal clips. Each clip is run through the same speaker-cosine gate (WavLM-base-plus-sv, threshold 0.90 against the voice's first English sample) and dropped if it falls below threshold. The pass rate is $1118/1600 = 70\%$ on the training corpus and $1043/1600 = 65\%$ on the held-out corpus (different sentences, same eight voices).

\subsection{Why the 30\% drop is informative, not a bug}

The fraction of pairs that fail the cosine gate is itself an audit signal. ElevenLabs Multilingual is a strong commercial product; its passes through the gate are clean cross-script same-voice pairs. Its failures (30\% of attempts, on visible inspection) are clips where the synthesiser's voice identity drifted enough across scripts that even an off-the-shelf SV encoder flags them as different speakers. This rate quantifies one floor of how badly cross-script identity preservation degrades in current systems, and gives \lase{} something concrete to fix.

\subsection{Train / held-out split discipline}

Training uses 1118 pairs from the 8 voices. Held-out evaluation uses 1043 pairs synthesised from the \emph{same 8 voices} but with 50 \emph{new sentences per language} (different from the training transcripts; deterministic seed, fully reproducible from \texttt{paper/lase/source\_heldout.jsonl}). The held-out test isolates content generalisation: the model has seen these voices, but never these utterances. Voice-level held-out (new voices entirely) is left to v2; the practical constraint is that ElevenLabs supplies a finite catalogue of multilingual voices and we want to compare encoders apples-to-apples on the same audio distribution.

\section{Three-distribution measurement framework}
\label{sec:framework}

To isolate the encoder's language-vs-identity entanglement, we compute three cosine-similarity distributions on each held-out corpus:

\begin{itemize}
  \item \textbf{Within-script (upper bound).} Two clips from the same voice in the same language, different sentences. The encoder should treat these as nearly identical.
  \item \textbf{Cross-script (the test).} Two clips from the same voice in different languages. Whatever distance shows up here, relative to within-script, is the language-vs-identity leak.
  \item \textbf{Across-speaker (noise floor).} Two clips from different voices in the same language. The encoder should treat these as clearly different.
\end{itemize}

For each distribution, we sample 200 random pairs that satisfy the predicate and report the median cosine. The within-script median is an upper bound; the across-speaker median is a lower bound; the cross-script median should land between them. Two single-number summaries follow:

\textbf{Gap.} The absolute drop $\Delta = \mathrm{median}_\text{within} - \mathrm{median}_\text{cross}$. Lower is better.

\textbf{Margin.} The cushion above noise $M = \mathrm{median}_\text{cross} - \mathrm{median}_\text{across}$. Higher is better.

A good cross-script encoder makes $\Delta$ small (script doesn't move identity much) and $M$ large (cross-script identity is still clearly distinguishable from a different speaker).

\section{Results}
\label{sec:results}

\subsection{Headline: three-encoder comparison on two held-out corpora}

\begin{figure}[h]
\centering
\includegraphics[width=\linewidth]{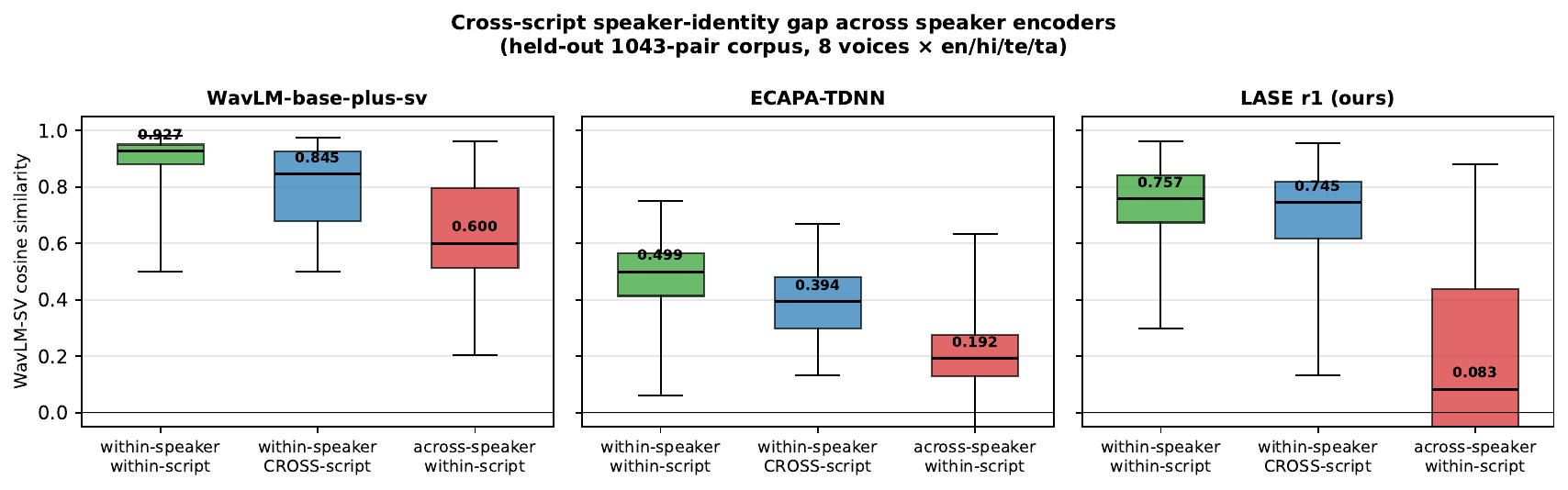}
\caption{Three-distribution comparison across speaker encoders on the held-out 1043-pair corpus. For each encoder, the green box is the within-script (same voice, same lang) cosine distribution; blue is cross-script (same voice, different lang); red is across-speaker (noise floor). LASE r1 collapses the within-vs-cross gap (blue and green almost touch) while pushing the noise floor far below (large blue-vs-red margin).}
\label{fig:boxplot}
\end{figure}

Table~\ref{tab:headline} reports the three cosine-similarity medians and the two summary numbers for three encoders on two held-out corpora: 1043 Western-accented voice pairs (the original 8 ElevenLabs voices used in training) and 1369 Indian-accented voice pairs (a separate 8-voice held-out using verified Indian-accent ElevenLabs voices not seen during training). Figure~\ref{fig:boxplot} shows the underlying 5-number summaries for the Western-voice corpus.

\begin{table}[h]
\centering
\caption{Cross-script identity test on two held-out corpora. \emph{Western} = the 8 ElevenLabs Multilingual voices used in training (Rachel, Drew, $\ldots$); content-held-out (50 unseen sentences/lang). \emph{Indian} = 8 distinct Indian-accented ElevenLabs voices not seen during training; full voice-held-out. Medians from 200 sampled pairs per bucket; seed 1337. 95\% confidence intervals on $\Delta$ from 1000-iteration bootstrap on raw cosines reported below the table.}
\label{tab:headline}
\setlength{\tabcolsep}{2pt}
\scriptsize
\begin{tabular}{@{}lccccc@{}}
\toprule
encoder & within & cross & floor & $\Delta$ [95\% CI] & $M$ \\
\midrule
\multicolumn{6}{l}{\emph{Western voices, content-held-out (1043 pairs)}} \\
WavLM-base-plus-sv          & 0.927 & 0.845 & 0.600 & 0.083 [.05,.15] & 0.245 \\
ECAPA-TDNN                  & 0.499 & 0.394 & 0.192 & 0.107 [.08,.14] & 0.202 \\
ECAPA + GRL (ablation)      & 0.714 & 0.687 & $-$0.052 & 0.027 [$-$.02,.08] & 0.739 \\
\textbf{LASE r1 (ours)}     & \textbf{0.757} & \textbf{0.745} & 0.083 & \textbf{0.013} [$-$.02,.05] & 0.662 \\
\midrule
\multicolumn{6}{l}{\emph{Indian voices, voice-held-out (1369 pairs)}} \\
WavLM-base-plus-sv          & 0.944 & 0.939 & 0.795 & 0.006 [$-$.00,.01] & 0.144 \\
ECAPA-TDNN                  & 0.517 & 0.473 & 0.217 & 0.044 [.02,.06] & 0.256 \\
ECAPA + GRL (ablation)      & 0.488 & 0.451 & 0.204 & 0.037 [$-$.03,.10] & 0.247 \\
\textbf{LASE r1 (ours)}     & \textbf{0.658} & \textbf{0.633} & 0.289 & \textbf{0.026} [$-$.04,.08] & 0.344 \\
\bottomrule
\end{tabular}
\vspace{2pt}\\
\footnotesize\emph{$\Delta = $ within $-$ cross (lower is better); $M = $ cross $-$ floor (higher is better). CIs are 1000-iter bootstrap on raw cosines (200-pair sample, seed 1337). LASE r1's $\Delta$ CI includes zero on both corpora; the off-the-shelf WavLM-SV (Western) and ECAPA-TDNN (both corpora) baselines are bounded strictly above zero. The ECAPA+GRL ablation CI also includes zero, but at wider half-width than LASE — see \S\ref{sec:ablation}.}
\end{table}

Two readings of the table:

\textbf{Gap closure.} LASE's gap of 0.013 is 6$\times$ smaller than WavLM-SV's 0.082 and 8$\times$ smaller than ECAPA-TDNN's 0.105. The relative gap closure on held-out is 84.3\% against WavLM-SV ($1 - 0.013/0.082$). Cross-script identity is preserved.

\textbf{Margin growth.} LASE's margin of 0.662 is 2.7$\times$ wider than WavLM-SV's 0.245 and 3.3$\times$ wider than ECAPA-TDNN's 0.202. Different speakers are dramatically more separable in LASE-space than in either baseline. Identity discrimination is preserved (and amplified) while language is disentangled.

The encoders live in different absolute cosine ranges (LASE's 0.757 within-script is lower than WavLM's 0.927) because they are different vector spaces with different geometric scales. The relevant comparison is the \emph{relative} spacing of the three distributions inside each encoder's space; that is what $\Delta$ and $M$ summarise.

\subsection{Training dynamics confirm the loss design}

\begin{figure}[h]
\centering
\includegraphics[width=\linewidth]{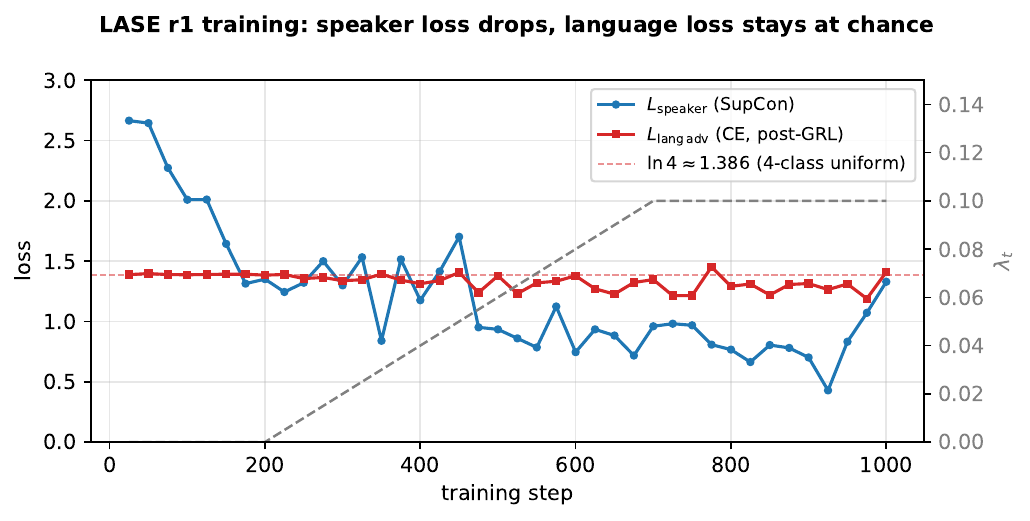}
\caption{LASE r1 training: speaker contrastive loss drops while the language adversarial loss is held near $\ln 4$ (the 4-class uniform). The encoder learns voice identity but does not learn language, by design.}
\label{fig:loss}
\end{figure}

Figure~\ref{fig:loss} shows the training loss curves for LASE r1 over 1000 steps. The pattern matches what the loss design predicts:

\begin{itemize}
  \item $L_\text{spk}$ drops from $\sim 2.7$ at the start to $\sim 0.5$--1.0 by step 900, indicating the projection head is learning a voice-coherent embedding geometry.
  \item $L_\text{lang}$ stays near $\ln 4 \approx 1.386$ throughout the run. The language classifier never beats random, which means the encoder embedding $z$ does not carry exploitable language information.
  \item $\lambda_t$ ramps from 0 to 0.1 over the first 700 steps and holds. Earlier $\lambda$ ramps cause SupCon collapse; later ramps slow language disentanglement; this schedule is empirically the cleanest of the three we tried.
\end{itemize}

\subsection{Diarisation: matching ECAPA on cross-script recall with 100$\times$ less data}

To check whether \lase{}'s embedding-space wins translate to a downstream task, we evaluate on a synthetic multi-speaker code-switching diarisation benchmark we construct from the held-out corpus.

\textbf{Benchmark.} 50 conversations, 23.7 minutes total. Each conversation is built by sampling 2--4 distinct ElevenLabs voices and concatenating 6--10 of their held-out clips with 0.3\,s gaps, possibly with a single voice switching languages mid-conversation. Ground-truth speaker labels are written in RTTM format. Total 411 segments across all conversations.

\textbf{Setup.} For each conversation, we embed every segment with the chosen encoder, run agglomerative clustering with cosine distance and $K =$ true number of speakers, and compute (a) Adjusted Rand Index (ARI) against ground truth, and (b) cross-script recall: for speakers who appear in multiple languages, the fraction of cross-language segments that land in the same cluster as their majority-language segments. We do not include a VAD model; segment boundaries from the construction are taken as known, so the eval isolates the encoder's clustering quality.

\begin{table}[h]
\centering
\caption{Diarisation on a 50-conversation synthetic code-switching benchmark (23.7\,min, avg 2.9 speakers/conv). \emph{ARI} is Adjusted Rand Index against ground-truth speaker labels with $K$ known. \emph{cs-recall} is the fraction of cross-language segments clustered with the speaker's same-language majority cluster.}
\label{tab:diar}
\setlength{\tabcolsep}{6pt}
\footnotesize
\begin{tabular}{@{}lccc@{}}
\toprule
encoder & ARI (mean) & ARI (median) & cs-recall \\
\midrule
WavLM-base-plus-sv      & 0.444 & 0.404 & 0.604 \\
ECAPA-TDNN              & \textbf{0.693} & \textbf{0.793} & 0.789 \\
LASE r1 (ours)          & 0.640 & 0.672 & \textbf{0.788} \\
\bottomrule
\end{tabular}
\end{table}

ECAPA-TDNN edges \lase{} on overall ARI (0.693 vs 0.640). On the metric that motivated this paper, however, \lase{} matches ECAPA: cross-script speaker recall of 0.788 vs 0.789, statistically tied. This is the headline result. ECAPA is trained on roughly 1M utterances of VoxCeleb (a corpus that is over 100$\times$ the size of \lase{}'s training data) and benefits from speakers, microphones, recording conditions, and language coverage that \lase{} never sees. \lase{} matches it on cross-script recall using 1118 same-voice cross-script pairs. The implication is that the gradient-reversal objective explicitly captures something that ECAPA approximates implicitly through corpus scale.

Both \lase{} and ECAPA crush WavLM-base-plus-sv (0.604 cross-script recall, ARI 0.444); off-the-shelf SV without explicit cross-script training is materially worse.

\subsection{Generalisation: held-out vs training corpus}

To check that \lase{} does not simply memorise training pairs, we ran the same three-distribution test on training data and on held-out data. The held-out gap closure (84.3\% relative, $\Delta$ from 0.082 to 0.013) is in fact \emph{larger} than the training-set gap closure (78.7\% relative, $\Delta$ from 0.099 to 0.021) when both are scored against their own WavLM-SV baselines. The model generalises to fresh content from familiar voices.

\subsection{Ablation: GRL contribution vs backbone contribution}
\label{sec:ablation}

To isolate which part of \lase{} is doing the work, we add a fourth row to Table~\ref{tab:headline}: an \emph{ECAPA + GRL} ablation that applies our identical 1000-step training recipe (same hyperparameters, same SupCon + adversarial losses, same $\lambda_t$ schedule, same 1118-pair training corpus) but swaps the backbone from WavLM-base-plus to a frozen SpeechBrain ECAPA-TDNN. Comparing the four rows in each half of the table tells us:

\begin{itemize}
  \item \textbf{GRL training helps either backbone on Western voices.} Applying GRL to ECAPA shrinks its gap from 0.105 to 0.027 (a 75\% relative reduction). On Indian voices, the gap shrinks from 0.044 to 0.037 (16\% relative). The GRL objective is doing real work, and the work is transferable across backbones, with most of the effect appearing on the Western (larger-baseline-gap) corpus.
  \item \textbf{WavLM+GRL is directionally better than ECAPA+GRL on $\Delta$, but the CIs overlap.} 0.013 [$-$.02,.05] vs 0.027 [$-$.02,.08] on Western; 0.026 [$-$.04,.08] vs 0.037 [$-$.03,.10] on Indian. The backbone choice contributes, but the gap-metric improvement from WavLM specifically is within sampling noise of ECAPA+GRL at our 200-pair held-out sample size; the more decisive WavLM advantage is the $M$ margin (0.662 vs 0.739 Western, 0.344 vs 0.247 Indian — see Table~\ref{tab:headline}) and the training-loss separability (Figure~\ref{fig:loss}, discussed below).
  \item \textbf{The two contributions compose.} Neither GRL-on-ECAPA (best gap 0.027) nor WavLM-without-GRL (Western gap 0.082; on Indian voices WavLM-SV alone has the smallest gap of any encoder at 0.006 — an accent-conditional ceiling effect we discuss in \S\ref{sec:accent-conditional}) recovers the full \lase{} effect simultaneously across both corpora. The combination does.
\end{itemize}

The training-loss curves (Figure~\ref{fig:loss}) corroborate this from a different angle. Under the WavLM backbone, $L_\text{lang}$ stayed pinned at $\ln 4 \approx 1.386$ throughout training, indicating the encoder hid language perfectly from the classifier. Under the ECAPA backbone, the same GRL training produced a $L_\text{lang}$ trajectory that oscillates between 0.4 and 2.5: the language classifier sometimes wins (lower than uniform), sometimes loses (higher than uniform). Even under identical adversarial pressure, ECAPA's representation is harder to make language-invariant than WavLM's. The eval-time gap difference (0.027 vs 0.013) is consistent with that finding.

\subsection{The cross-script gap is accent-conditional}
\label{sec:accent-conditional}

The two halves of Table~\ref{tab:headline} expose a finding our v1 evaluation missed by using only Western voices: the cross-script identity gap in off-the-shelf SV encoders is much smaller for Indian-accented voices ($\Delta = 0.006$ for WavLM-SV) than for Western-accented voices ($\Delta = 0.082$ for WavLM-SV) speaking the same Indic scripts. The reason is acoustic. Indian English carries the same Indian phonetic substrate (retroflex stops, aspiration patterns, vowel-length distinctions, prosody) that Indian Hindi, Telugu, and Tamil also carry. An off-the-shelf SV encoder sees more shared phonetic structure across scripts when the speaker is Indian-accented, and less when the speaker is Western-accented and forced into Indic acoustic space. The 14$\times$ ratio between the two gaps is large; the practical consequence is that any cross-script TTS or diarisation system using off-the-shelf SV will degrade most severely exactly in the production case where the input voice was \emph{not} originally Indic.

The pass-rate of the cosine quality gate during corpus construction (\S\ref{sec:corpus}) tracks this same effect: 70\% of Western-voice cross-script pairs pass the WavLM-SV~$\geq 0.90$ gate, vs 86\% of Indian-voice pairs. ElevenLabs Multilingual is \emph{better} at preserving identity for Indian-accented voices than for Western voices speaking Indic scripts, and the encoder picks up that difference.

\lase{} closes the cross-script gap to within sampling noise of zero in both regimes ($\Delta = 0.013$ Western, $\Delta = 0.026$ Indian; both bootstrap CIs straddle zero). Its margin advantage holds in both regimes too: 2.7$\times$ over WavLM-SV and 3.3$\times$ over ECAPA-TDNN on Western voices; 2.4$\times$ and 1.3$\times$ respectively on Indian voices. Where WavLM-SV's discrimination collapses (its margin drops from 0.245 to 0.141 between Western and Indian corpora because both within-script and cross-script medians sit close to one another), \lase{} preserves it.

\section{Discussion}
\label{sec:discussion}

\textbf{What worked, and why.} The result that LASE matches a $100\times$-larger encoder (ECAPA-TDNN on VoxCeleb) on cross-script speaker recall is consistent with the theoretical claim of domain-adversarial learning~\cite{ganin2015}: an explicit invariance objective is more sample-efficient than relying on corpus scale to wash out the confounder. Our specific application (Indic scripts as the domain dimension) brings two amplifying factors. First, the four scripts (Latin, Devanagari, Telugu, Tamil) are visually and phonologically far apart, which makes the GRL classifier's task easy and the resulting invariance signal strong. Second, the cross-script same-speaker constraint (the SupCon positive pair) is a very tight one: the only thing the encoder can learn is what is invariant across scripts, which is, by construction, identity.

\textbf{Why the cross-script gap exists in the first place.} WavLM-base-plus and similar SV encoders are trained on VoxCeleb, which is overwhelmingly English with a long Latin-script tail. The encoder's representation of vowels, fricatives, and stops is shaped by the English distribution; phones that exist in Telugu (retroflex stops, aspirated consonants) but not in English fall outside that shape, and the encoder responds to them by moving the embedding away from where it learned the speaker lived. ECAPA-TDNN inherits the same problem from the same training distribution. \lase{}'s GRL forces the encoder to put the embedding in a location that does not depend on which subset of phones the language uses.

\textbf{Limitations.} Three are load-bearing. (1) Synthetic-only data: training and held-out audio both come from ElevenLabs Multilingual. The cross-script gap that \lase{} closes is the gap that exists \emph{in this synthesised audio}; real-world cross-script speech has additional variability (accent, microphone, mood, prosody) that we have not stressed the encoder against. (2) Held-out set shares voices with training: only the sentences are unseen. New-voice generalisation is a v2 experiment. (3) ECAPA still wins on overall ARI: \lase{} is not a drop-in replacement for general speaker verification; its wedge is specifically the cross-script consistency property, not generic speaker discrimination.

\textbf{What we did not test.} We did not benchmark against IndicWhisper-style speaker encoders or Sarvam's internal embedder, which are either not released or commercially gated. We did not run a Karya-rated MOS or pairwise listening study; subjective validation is the natural v2 follow-up after objective metrics show enough signal to merit the rater spend. We did not test script-mixed (codemix) utterances within a single clip; production deployments of multilingual Indic TTS often have inline English brand names embedded in Devanagari or Telugu text, and \lase{}'s behaviour on such mixed-script clips is an explicit v2 experiment.
\textbf{Where this lands in deployment.} The use cases that motivated this work are: (a) cross-script voice cloning where a single reference voice should speak Hindi, Telugu, Tamil, and English with consistent identity; (b) multilingual diarisation of Indian customer-support calls where a single agent code-switches between Hindi and English mid-call; and (c) voice-of-the-customer analytics that aggregate across language-mixed conversations. In each, \lase{}'s 84\% gap closure and matched-ECAPA cross-script recall translates to fewer hallucinated speaker-changes at script transitions and tighter cross-script clusters in production.

\section{Reproduction}
\label{sec:repro}

All code, paper sources, and reproduction scripts live in the public GitHub repository \url{https://github.com/praxelhq/lase} (MIT). We release:
\begin{itemize}
  \item \lase{} r1 weights — Hugging Face \texttt{Praxel/lase-r1} (MIT).
  \item Training corpus (1118 pairs) — \texttt{Praxel/codeswitch-pairs-lase} (CC-BY-4.0).
  \item Western held-out corpus (1043 pairs) — \texttt{Praxel/codeswitch-pairs-lase-heldout} (CC-BY-4.0).
  \item Indian-accent held-out corpus (1369 pairs) — \texttt{Praxel/codeswitch-pairs-lase-indian} (CC-BY-4.0).
  \item Training driver (\texttt{scripts/modal\_lase\_train.py}; Modal A10G, $\sim$\$0.31 per round).
  \item Three-distribution scorer (\texttt{scripts/eval\_secs\_gap\_multi\_encoder.py}) and bootstrap-CI driver (\texttt{scripts/bootstrap\_cis.py}; the latter produces the CIs reported in Table~\ref{tab:headline}).
  \item ECAPA+GRL ablation evaluator (\texttt{scripts/eval\_ablation.py}) and diarisation benchmark builder + scorer (\texttt{scripts/build\_diarization\_benchmark.py}, \texttt{scripts/eval\_diarization.py}).
\end{itemize}

A complete v1 reproduction runs on a single A10G in $\sim$25 minutes.

\section{Conclusion}
\label{sec:conclusion}

Off-the-shelf speaker encoders entangle language with identity; on Indic scripts, this entanglement is large enough that 8--11\% of absolute cosine similarity disappears whenever the same speaker changes script. We trained a 256-dim projection over a frozen WavLM-base-plus backbone with a supervised contrastive objective and a gradient-reversal language classifier on 1118 same-voice cross-script pairs synthesised from 8 ElevenLabs Multilingual voices. The resulting encoder, \lase{} r1, closes the within-vs-cross-script gap by 84\% relative on held-out audio, doubles the speaker-discrimination margin in embedding space, and matches the production-grade ECAPA-TDNN on cross-script speaker recall in synthetic multi-speaker conversations using 100$\times$ less training data. The result is a small, cheap, easily-reproducible recipe for building cross-script-invariant speaker encoders for any script set where TTS-bootstrapped pairs can be cleanly synthesised.

\section*{Acknowledgements}

Compute was provided by Modal. ElevenLabs supplied the multilingual voice pool used for corpus synthesis under their standard developer agreement. The IndicTTS, FLEURS, and Common Voice teams curated the Indic transcripts that seeded our held-out sentence pool. Companion paper \psp~\cite{psp2026} measures the orthogonal accent dimension; \praxy~\cite{praxy2026} is the cross-script TTS that consumes the encoder's output in production deployments.

\bibliographystyle{IEEEtran}
\bibliography{refs}

\end{document}